\documentclass[twocolumn,superscriptaddress,aps,floatfix]{revtex4}
\usepackage{amsmath}
\usepackage{amssymb}
\usepackage{graphicx}
\usepackage{color}
\usepackage{physics}

\begin{document}

\title{Localized Singlets and Ferromagnetic Fluctuations in the Dilute Magnetic Topological Insulator Sn$_{0.95}$Mn$_{0.05}$Te}

\author{D.~Vaknin}
\affiliation{Ames Laboratory, Ames, IA, 50011, USA}
\affiliation{Department of Physics and Astronomy, Iowa State University, Ames, IA, 50011, USA}

\author{Santanu Pakhira}
\affiliation{Ames Laboratory, Ames, IA, 50011, USA}

\author{D. Schlagel}
\affiliation{Ames Laboratory, Ames, IA, 50011, USA}

\author{F.~Islam}
\affiliation{Department of Physics and Astronomy, Iowa State University, Ames, IA, 50011, USA}

\author{Jianhua Zhang}
\affiliation{Department of Physics and Astronomy, Iowa State University, Ames, IA, 50011, USA}

\author{D.~Pajerowski}
\affiliation{Oak Ridge National Laboratory, Oak Ridge, TN, 37831, USA}

\author{C.~Z.~Wang}
\affiliation{Ames Laboratory, Ames, IA, 50011, USA}

\author{D.~C.~Johnston}
\affiliation{Ames Laboratory, Ames, IA, 50011, USA}
\affiliation{Department of Physics and Astronomy, Iowa State University, Ames, IA, 50011, USA}

\author{R.~J.~McQueeney}
\affiliation{Ames Laboratory, Ames, IA, 50011, USA}
\affiliation{Department of Physics and Astronomy, Iowa State University, Ames, IA, 50011, USA}

 \date{\today}

\begin{abstract}
The development of long-range ferromagnetic (FM) order in dilute magnetic topological insulators can induce dissipationless electronic surface transport via the quantum anomalous Hall effect.  We measure the magnetic excitations in a prototypical magnetic topological crystalline insulator, Sn$_{0.95}$Mn$_{0.05}$Te, using inelastic neutron scattering.  Neutron diffraction and magnetization data indicate that our Sn$_{0.95}$Mn$_{0.05}$Te sample has no FM long-range order above a temperature of 2 K.  However, we observe slow, collective FM fluctuations ($<$~70 $\mu$eV), indicating proximity to FM order.  We also find a series of sharp peaks originating from local excitations of antiferromagnetically (AF) coupled and isolated Mn-Mn dimers with $J_{\rm AF}=460$~$\mu$eV\@.  The simultaneous presence of collective and localized components in the magnetic spectra highlight different roles for substituted Mn ions, with competition between FM order and the formation of AF-coupled Mn-Mn dimers.
 \end{abstract}

\maketitle
Magnetic topological insulators belong to a promising class of materials that can host novel surface transport phenomena \cite{Tokura19}. For example, the quantum anomalous Hall effect has been demonstrated by inducing long-range ferromagnetic (FM) order via the addition of small concentrations of magnetic ions into tetradymite topological insulators \cite{Zhang13, Chang13, Chang15}.  Similar to dilute FM semiconductors, further development of dilute magnetic topological systems requires a deep understanding of the microscopic origin of the magnetic interactions which is highly dependent on chemical disorder and/or inhomogeneity.

Inelastic neutron scattering (INS) is a powerful method to resolve the energy scales of the magnetic interactions that give rise to bulk magnetism.  In these dilute magnetic systems,  both short-range exchange coupling and long-range Ruderman-Kittel-Kasuya-Yosida (RKKY) interactions \cite{RKKY} mediated by conduction electrons are present.  The distribution of magnetic ions and the competition between these interactions play a decisive role in the magnetic ground state and the efficacy of magnetic coupling to surface Dirac states.  

\begin{figure*}
\includegraphics[width=1.0\linewidth]{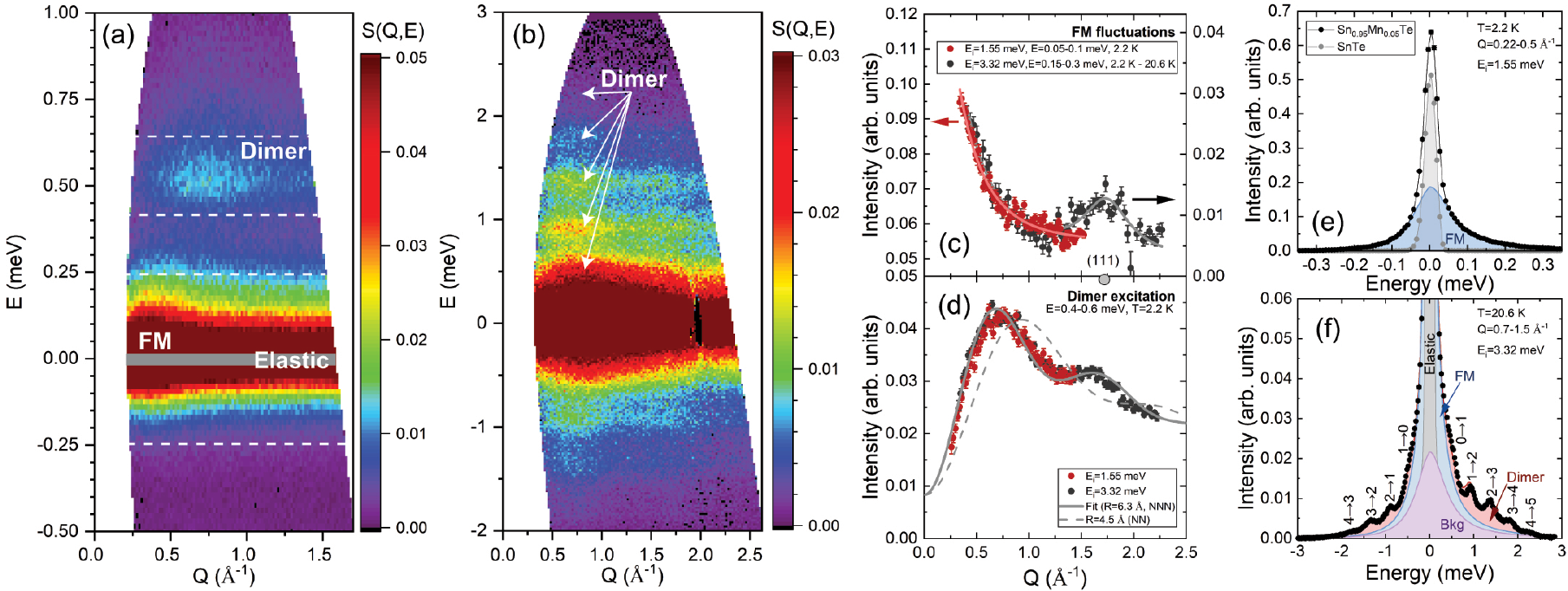}
\caption{\footnotesize Inelastic neutron scattering intensities from Sn$_{0.95}$Mn$_{0.05}$Te measured at (a) $T=$ 2.2 K and $E_{i}=$ 1.55 meV and (b) $T=$ 20.6~K and $E_{i}=$ 3.32 meV\@.   In (a), the gray band at $E=0$ represents the elastic resolution FWHM of 0.036 meV.  (c) The $Q$-dependence of the low energy FM fluctuations for both $E_{i}=$1.55 and 3.32 meV\@.  Lines correspond to Lorentzian fits. (d) The $Q$-dependence of the dimer excitation at 0.5 meV for both $E_{i}=$1.55 and 3.32 meV at $T=$ 2.2~K\@. The solid line is a fit to the dimer form factor and the dashed line is a calculation of the dimer form factor for nearest-neighbors with $R_{\rm{NN}}=$ 4.5 \AA. (e) The low-$Q$ spectra of SnTe and Sn$_{0.95}$Mn$_{0.05}$Te consisting of elastic (gray shaded area) and quasielastic FM fluctuations (blue shade), respectively.  (f) The high-$Q$ spectrum at $T= 20.6$~K consisting of elastic (gray shade), FM (blue shade), dimer (red shade), and background (purple shade) contributions.  Lines and shaded regions are fits as described in the text. Except for panel (e), data from the parent SnTe compound has been subtracted.}
\label{fig1}
\end{figure*}

SnTe is a IV-VI semiconductor with a simple rock-salt structure that possesses a narrow band gap at four equivalent points (the $L$-points) in the Brillouin zone which are related to each other by mirror symmetry \cite{Hsieh12}.  By virtue of its non-trivial inverted band topology, SnTe is demonstrated to be a topological crystalline insulator that exhibits metallic surface Dirac cones hosting high-mobility chiral electrons \cite{Tanaka12, Yu12}.  These states are topologically protected by the mirror symmetry of the crystal with respect to \{110\} mirror planes \cite{Hsieh12, Yang14, Wang16}. For FM order with magnetic moments aligned in the mirror plane, broken time-reversal symmetry can give rise to the quantum anomalous Hall effect \cite{Wang16}. In SnTe, it has been shown that FM order can be stabilized by substituting small concentrations of magnetic Mn$^{2+}$ ions for Sn above a threshold concentration $x>0.03$ in Sn$_{1-x}$Mn$_{x}$Te \cite{Inoue77, Eggenkamp94}.

In this Letter, we report INS investigations of the magnetic dynamics in a prototypical dilute magnetic topological insulator Sn$_{0.95}$Mn$_{0.05}$Te that does not exhibit long-range FM order above 2 K.  We observe collective FM excitations mediated by long-range interactions and well-defined local excitations of antiferromagnetically (AF)-coupled Mn-Mn dimers. Our analyses are consistent with random Mn substitution on the Sn sites and highlight the competition between long- and short-range magnetic interactions. The AF dimers occur between next-nearest-neighbor (NNN) Mn spins as supported by our {\it ab initio} electronic structure calculations, instead of between nearest-neighbor (NN) Mn spins as might have been anticipated.

The powder samples of SnTe and Sn$_{0.95}$Mn$_{0.05}$Te used for this study were synthesized using solid-state reaction from stoichiometric quantities of Sn, Mn, and Te as described in the Supplemental Material (SM) \cite{SM}. Analysis of scanning electron microscopy and electron dispersive microscopy show the sample to be single phase with a composition close to the nominal values and x-ray powder diffraction measurements confirm the SnTe structure.  Magnetization measurements were carried out using a Quantum Design MPMS magnetometer.  Details of sample growth and characterization are given in the SM \cite{SM}.  

INS measurements were performed on the Cold Neutron Chopper Spectrometer (CNCS) at the Spallation Neutron Source at Oak Ridge National Laboratory. Sixteen grams of each sample were loaded into a 1/2-inch diameter aluminum can and attached to a closed-cycle refrigerator for measurements at $T=2.2$ to 20.6~K using incident neutron energies of $E_{i}=$ 1.55 and 3.32 meV\@.  Our magnetization and neutron diffraction data indicate that our $x=0.05$ sample has no long-range FM order, highlighting known sensitivities of the magnetic ordering transition in Sn$_{1-x}$Mn$_{x}$Te to both the Mn concentration and also the carrier concentration \cite{Story86, Vennix93, Eggenkamp94}. 

Figure~1(a) shows INS data from Sn$_{0.95}$Mn$_{0.05}$Te taken at $T=$ 2.2 K and $E_{i}=$ 1.55 meV as a function of momentum transfer ($Q$) and energy transfer ($E$). Measurements of the parent SnTe compound have been subtracted in order to remove phonon and other nonmagnetic background contributions. Two main excitation features are observed; quasielastic (QE) fluctuations near $E\approx0$ and an inelastic transition near 0.5 meV.  At a higher $T=$ 20.6 K and a larger $E_{i}=$ 3.32 meV, Fig.~2(b) clearly shows a band of equally-spaced excitations in addition to the 0.5 meV excitation.  

The origin of these two different contributions to the scattering can be ascertained from the $Q$-dependencies. With respect to the QE fluctuations, Fig.~1(c) shows that the intensity of the narrow QE response at 2.2 K and low energies of $E=$ 0.05--0.1 meV increases as $Q\rightarrow0$.  Also, Fig.~2(c) shows a subtraction of the 20.6 K data from the 2.2 K data with $E_{i}=$ 3.32 meV, clearly revealing that the QE fluctuations peak at nuclear Bragg centers $Q=0$ and $Q$(111)= 1.7~\AA$^{-1}$.  Based on these observations, we ascribe the QE signal to FM fluctuations.  For the inelastic excitations, Fig.~2(d) shows a $Q$-cut obtained from integrating over the 0.5 meV excitation from $E=$ 0.4 to 0.6 meV.  The cut shows an oscillatory $Q$-dependence that goes to zero as $Q\rightarrow0$ which is characteristic of dimer scattering from pairs of Mn ions, as described below.  Inspection of Fig.~1(b) shows that the set of excitations observed at higher temperatures obey the same $Q$-dependence and can all be associated with Mn-Mn dimers.

Based on this information, the spectrum of magnetic excitations in Sn$_{0.95}$Mn$_{0.05}$Te can be broken down into contributions from localized dimer excitations, FM fluctuations, and a magnetic background contribution.  In principle, the background contains contributions from other magnetic configurations, such as lone spins, trimers, and larger clusters.  The three contributions to the inelastic spectrum are represented as
\begin{equation}
S(Q,E)=S_{\rm dimer}(Q,E)+S_{\rm FM}(Q,E)+S_{\rm bkg}(Q,E).
\end{equation}
 
We now describe the FM fluctuations, which presumably occur from the long-range RKKY coupling between non-dimerized Mn ions and can be represented by relaxational dynamics
\begin{equation}
S_{\rm FM}(Q,E>0) = f^{2}(Q)\frac{S_{\rm FM}(Q)\Gamma_{\rm FM}}{\Gamma_{\rm FM}^2+E^2}
\label{Eqn:FM}
\end{equation}
where $\Gamma_{\rm FM}$ is the relaxation rate and detailed balance is obeyed $S_{\rm FM}(Q,E<0)=S_{\rm FM}(Q,E>0)e^{-|E|/k_{\rm B}T}$, where $k_{\rm B}$ is Boltzmann's constant.  The $Q$-dependence is a Lorentzian [$S_{\rm FM}(Q)=\chi_{0}/(Q^{2}+\kappa^{2})$] centered at nuclear Bragg peaks with a half-width ($\kappa$) corresponding to the inverse FM correlation length. From fits shown in Fig.~1(c), we find $\kappa=0.27\pm0.06$ \AA$^{-1}$, corresponding to a correlation length of 3-4 unit cells.  Figure~1(e) shows high-resolution 1.55 meV data taken on both SnTe and Sn$_{0.95}$Mn$_{0.05}$Te at $T=$ 2.2 K that is sampled close to $Q=0$ where dimer scattering is weak.  The SnTe spectrum is entirely elastic whereas the QE contribution in the Mn-substituted sample is apparent.  Lorentzian fits to the relaxational QE spectrum were performed at several temperatures with results shown in Fig.~2(a). From the fit in Fig.~1(e), we obtain $\Gamma_{\rm FM}=68 \pm 1$ $\mu$eV at 2.2~K.

We now turn to analysis of the dimer contribution to the scattering, which enables the determination of the short-range exchange coupling within the Mn-Mn dimers.  This contribution is obtained by assuming that a Heisenberg Hamiltonian
\begin{equation}
H=J\mathbf{s}_{1}\cdot \mathbf{s}_{2} =\frac{J}{2}[S(S+1)-2s(s+1)]
\label{Heisen}
\end{equation}
is applicable for the dimers. Here $\mathbf{s}_{1}$ and $\mathbf{s}_{2}$ are the operators for individual Mn spins of magnitude $s$ and $S$ is the total spin quantum number of the dimer.  For Mn$^{2+}$, individual spins have magnitude $s=$ 5/2 and $S$ can take on integer values from 0 to 5. The dimer energy levels are given by $\mathcal{E}(S)=\frac{J}{2}[S(S+1)-2s(s+1)]$. 

The cross-section for neutron scattering from dimers is given by \cite{Furrer13}
\begin{equation}
\begin{aligned}
S_{\rm dimer}(Q,E) = Af^{2}(Q)\Big[1+(-1)^{\Delta S}\frac{\mathrm{sin}(QR)}{QR}\Big]\times \\
\frac{e^{-\mathcal{E}(S)/k_{\rm B}T}}{Z}\abs{\matrixel{S}{\hat{T}}{S'}}^{2}\delta[E+\mathcal{E}(S)-\mathcal{E}(S')]
\label{xs}
\end{aligned}
\end{equation}
where $A$ is a scale factor, $f(Q)$ is the single-ion Mn magnetic form factor, $R$ is the Mn-Mn dimer distance, $Z$ is the partition function for the dimer levels, and $\Delta S = S'-S$ is the change in the total spin quantum number. Neutron scattering can observe both QE fluctuations ($\Delta S=0$) of the total dimer spin and transitions between dimer states with $\Delta S= \pm 1$. Dimer-state transitions $S \rightarrow S+1$ form a set of five equally-spaced excitations occurring at $\mathcal{E}(S+1)-\mathcal{E}(S)=J(S+1)$. The matrix elements contribute to the neutron intensities and are given by
\begin{equation}
\abs{\matrixel{S}{\hat{T}}{S'}}^{2} \propto (2S+1)(2S'+1)\begin{Bmatrix} S&S'&1 \\ s&s&s \end{Bmatrix}
\end{equation}
where the quantity in the curly braces is the Wigner 6-$j$ symbol.

We first note that the $Q$-dependence for dimer-state transitions with $\Delta S= \pm 1$ follows the dimer form factor $ f^{2}(Q)\Big[1-\frac{\mathrm{sin}(QR)}{QR}$\Big] as shown in Fig.~1(d). Interestingly, we find a Mn-Mn dimer distance of $R=$ 6.3(2) \AA\ that corresponds to \textit{next}-nearest-neighbor (NNN) pairs on the Sn/Mn sublattice, and not NN where $R=4.5$~\AA~(as shown in Fig.~1(d) for comparison). We come back to this point below when we discuss results from density-functional theory calculations.

To determine the dimer contribution to the total magnetic spectrum, we analyzed the coarser resolution 3.32 meV data over a $Q$-range from 0.7--1.2 \AA$^{-1}$. For these fits, we fixed $S_{\rm FM}$ [Eqn.~(\ref{Eqn:FM})] by scaling the parameters obtained from high-resolution fits described above to values appropriate for our sampling of the 3.32 meV data. The inelastic lines were fit to Lorentizans with half-width $\gamma$.  Fits of the remaining intensity to $S_{\rm dimer}$ were poor which required the introduction of an additional, broad QE response, $S_{\rm bkg}$. An example of a fit to the 20.6 K spectrum is shown in Fig.~1(f) where the full set of five spin-state transitions are observed.

Spectra were fit at several temperatures and we find that the Heisenberg dimer model parameters are essentially $T$-independent [$J=$ 0.460(2) meV and $\gamma=$ 0.142(6) meV] whereas the dimer scale factor $A$ and $S_{\rm bkg}$ parameters change substantially below $\sim$ 7 K.  Since $A$ is proportional to the number of Mn dimers, there is no expectation that it should be $T$-dependent.  Thus, we repeated the fits with $A$ fixed to its average value.  This action does not meaningfully change $J$ or $\gamma$ [Fig. 2(b)], but does change the temperature evolution of $S_{\rm bkg}$, as shown in Fig.~2(c).  We also note that the $S=0 \rightarrow1$ transition energy becomes progressively larger than $J$ at low-$T$, but other transitions are relatively unaffected.  The distortion of the equally-spaced dimer transition states indicates that other interactions are present beyond the isotropic Heisenberg model. To account for this, we added a phenomenological parameter that describes the shift of the ground state energy ($E_{0}$), as shown in Fig.~2(b).  Full fitting results are described in the SM \cite{SM}. 
\begin{figure}
\includegraphics[width=1.0\linewidth]{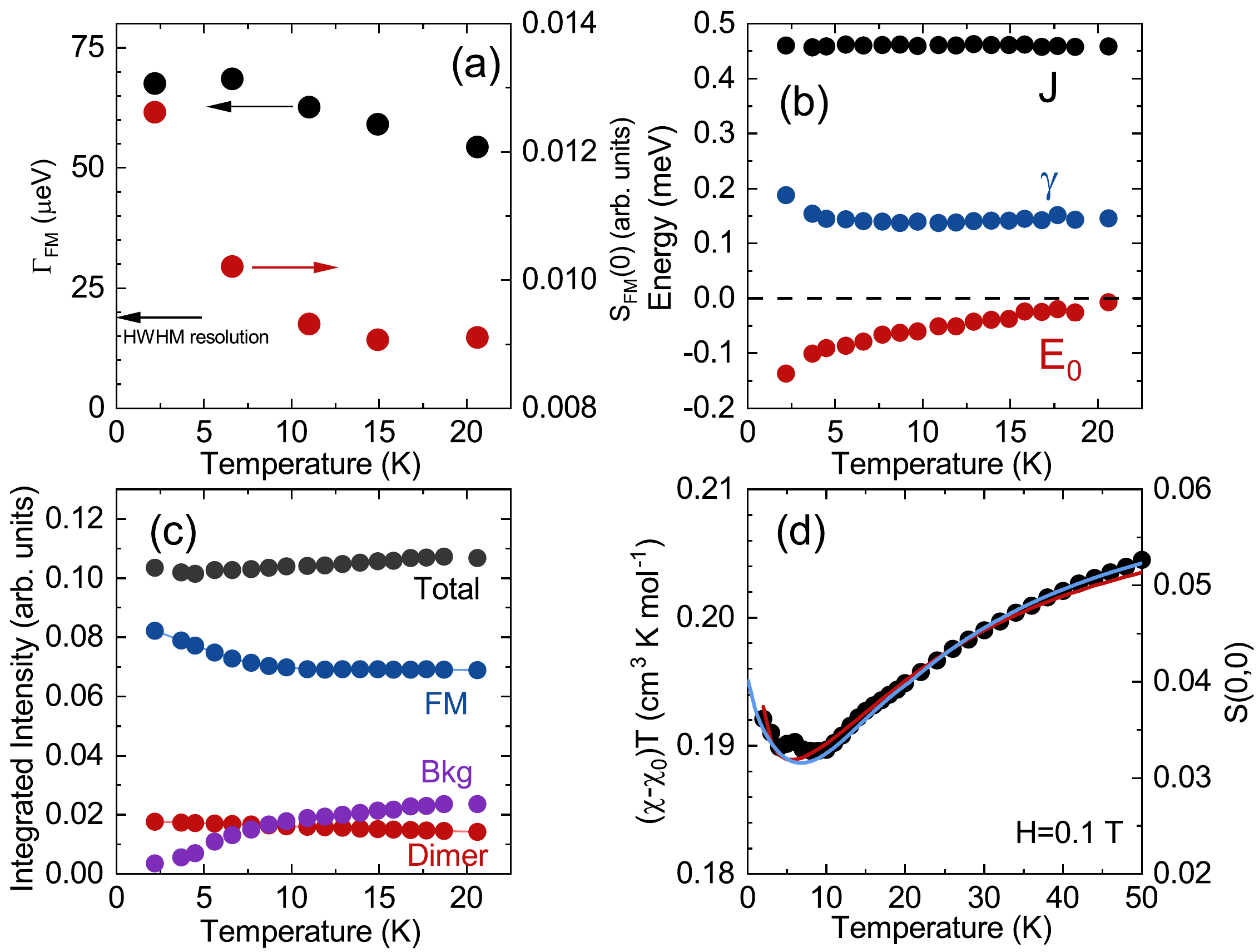}
\caption{\footnotesize (a) Amplitude $S_{\rm FM}(0)$ and relaxational width $\Gamma_{\rm FM}$ of the FM fluctuations as a function of temperature.  (b) Temperature dependence of dimer model parameters determined from fits to the 3.32 meV data.  (c) Comparison of the integrated intensities of the different contributions to the total magnetic neutron cross-section with $E_{i}=$ 3.32 meV. (d) The bulk susceptibility measured at $H = 0.1$~T and plotted as ($\chi-\chi_{0})T$ versus $T$ (black circles).  The red line is a fit to a dimer model and the blue line is the bulk susceptibility determined from neutron scattering data (to within a scale factor).  In (d), the INS calculations are barely distinguishable from the experimental data.}
\label{fig2}
\end{figure}

Detailed fits to the spectra reveal that Mn-Mn NNN dimers have strong AF exchange coupling as compared to the FM fluctuations ($J/\Gamma_{\rm FM}=6.6$) resulting in some fraction of the substituted Mn ions adopting a singlet ground state that cannot participate in FM fluctuations.  The fraction of Mn within these dimers ($f_{\rm dimer}$) can be estimated by integrating the intensity of the different components to the spectrum, as shown in Fig.~2(c).  The FM fluctuations are dominant, accounting for more than 65\% of the cross-section, and this fraction increases as the temperature is lowered, largely at the expense of the broad background contribution.  $f_{\rm dimer}$ is roughly constant and accounts for $\sim$17\% of all magnetic scattering.  Since we only fit dimer-state transitions, the slight decrease in the dimer intensity at higher temperatures is accounted for by the growth of dimer QE scattering ($\Delta S=0$) which was not explicitly accounted for in the fit, but presumably is subsumed in $S_{\mathrm{bkg}}$.  As expected, the total of all contributions to the magnetic intensity is nearly independent of $T$.

We also performed complementary bulk magnetic susceptibility measurements, as shown in Fig.~2(d).  The bulk susceptibility has a contribution from dimer states, which are represented as
\begin{equation}
\chi=\chi_{0}+(1-f_{\rm dimer})\chi_{\rm CW}+f_{\rm dimer}\chi_{\rm dimer}
\end{equation}
where$\chi_0$ is a constant term, $\chi_{\rm CW}=C_{\rm Mn}/(T-\theta_{p})$ is the non-dimer Mn susceptibility,  $C_{\rm Mn}=Ng^2\mu_{B}^{2}s(s+1)/3k_{\rm B}$.  Using the Heisenberg model [Eq.~(\ref{Heisen})], it can be shown that the dimer contribution is (see SM) \cite{SM}
\begin{equation}
\chi_{\rm dimer}=\frac{C_{\rm Mn}}{2s(s+1)}\frac{1}{ZT}\sum_{S}S(S+1)(2S+1)e^{-\mathcal{E}(S)/kT}.
\end{equation}
The formation of AF dimer singlets suppresses the low-temperature susceptibility and our fits find that $J=0.60(3)$ meV, $\theta_{\rm p}=$ 0.083(5)~K, $f_{\rm dimer}= 0.147(3)$, and $C_{\rm Mn}= 0.216(1)$\,cm$^{3}$\,K\,mol$^{-1}$. Both $J$ and $f_{\rm dimer}$ compare favorably to the INS results and the value of $C_{\rm Mn}$ corresponds to that expected for $x=0.05$.  Finally, the INS data can be compared directly to the bulk susceptibility according to the relation $(\chi-\chi_{0}) T \propto S_{\rm FM}(0,0)+S_{\rm dimer}(0,0)$, as shown in Fig.~2(d).  It is satisfying to find agreement between the INS and $\chi(T)$ data on the susceptibility suppression and even the upturn below 10~K caused by developing FM correlations.

The unusual observation that dimers appear only at NNN positions might suggest that Mn substitution is not truly random and strong repulsion occurs for NN Mn pairs.  To address this issue, we investigated the energetics of the formation of different Mn-Mn dimer pairs using density functional theory (DFT) with methods described in the SM\cite{SM}.  Because the anti-site and interstitial defects are highly energetically unfavorable \cite{Wang14}, here we only considered the substitutional defect. In the calculations, it was found that Mn atoms preferred to substitute for Sn atoms, consistent with previous experimental results \cite{Inoue81,Yeoh88} and calculations \cite{Wang16}.  We calculated the total energy of the cell for Mn pairs in either a FM or AF configuration, as shown in Fig.~3(a). The geometry optimizations were performed until all the forces acting on each atom are less than 0.01eV \AA$^{-1}$.

In seeming contrast to the INS observations, our DFT calculations find that NNN Mn pairs are least favored energetically; however, this energy difference ($\sim$ 450 K) is lower than the synthesis temperature of the samples (see SM \cite{SM}) suggesting random substitution.  A perhaps more surprising discovery is that the NNN Mn pairs have a much larger magnetic splitting than pairs at other distances, even NN pairs.  This arises from the linear Mn-Te-Mn bond configuration of the NNN dimer that strongly enhances AF superexchange as compared to the 90$^{\circ}$ bond for the NN dimer where superexchange is weaker.  Similar energetics can be found in transition metal oxides with the rock-salt structure, such as MnO \cite{Fischer09}.  The magnetic energy difference provides an estimate of the dimer exchange $E({\rm FM})-E({\rm AF})=2Js(s+1)$, as shown in Fig.~3(b).  For the NNN pair, we obtain $J=$ 1.4 meV which is somewhat larger, but of the same order of magnitude, as the experimentally-determined value.  Dimer pairs at other distances are only weakly coupled magnetically and therefore behave as free moments that participate in FM fluctuations within the temperature and energy scales studied. 

\begin{figure}
\includegraphics[width=1.0\linewidth]{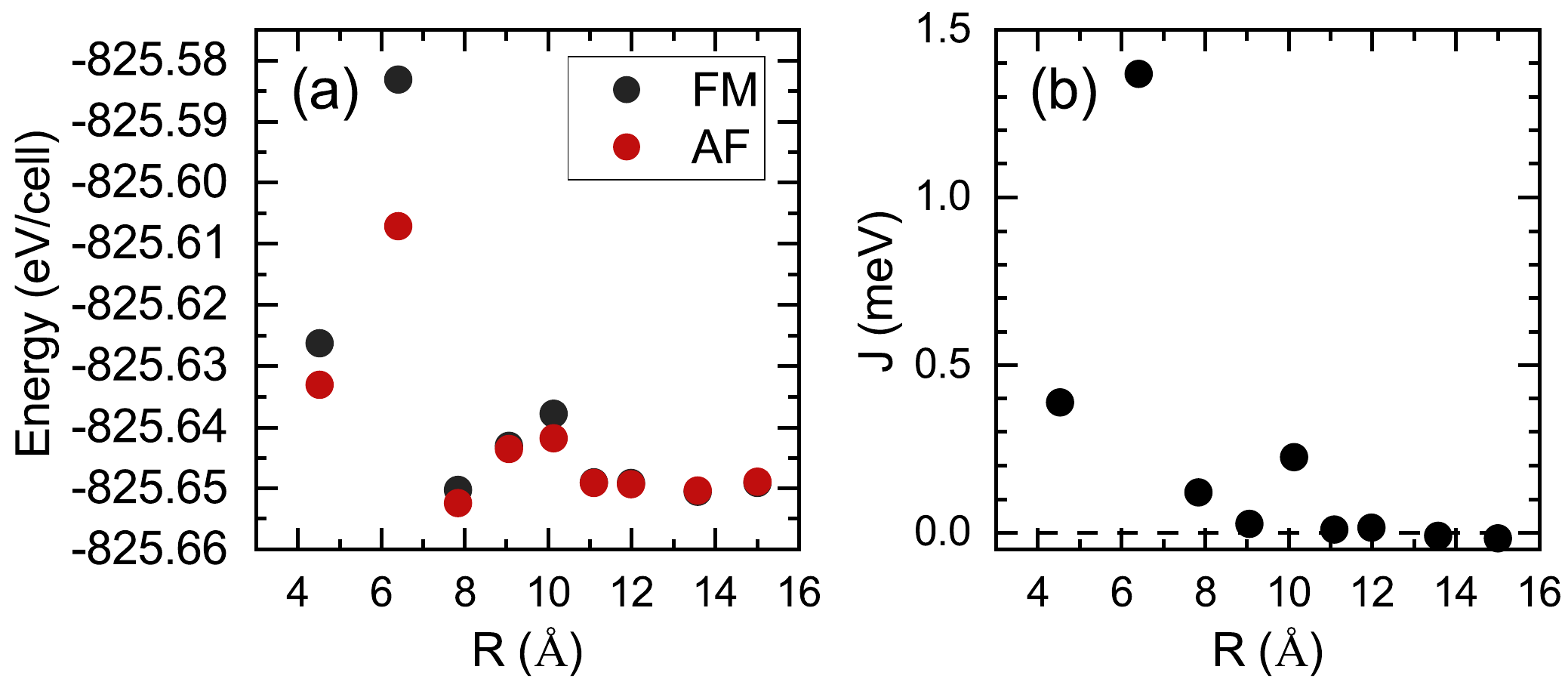}
\caption{\footnotesize (a) Density functional theory calculations of the total energy of a $3\times3\times3$ supercell with a single pair of Mn atoms at different distances in either a FM or AF configuration. (b) DFT estimate of the magnetic exchange energy for different Mn pair distances.}
\label{fig3}
\end{figure}

We are now in a position to compare experimental results for $f_{\rm dimer}$ to expectations for a random solid solution of Mn substituting for Sn on a FCC sublattice.  In our case, we assume that Mn NN pairs are present, but magnetically uncoupled.  In this special case, we are concerned with the probability (without regard for NN occupancies) that a substituted Mn ion is paired with one other Mn ion in a NNN dimer.  This probability is $f_{\rm dimer}=6x(1-x)^{10}$ which for $x=0.05$ gives $f_{\rm dimer}=0.18$.  Reasonable agreement of these numbers with both neutron and magnetization data are good confirmation that Mn substitutes randomly for Sn.

Overall our results provide two major findings.  The first is that Sn$_{1-x}$Mn$_{x}$Te is a dilute topological magnetic system where quenched disorder provides an ideal random alloy.  This is an important result, since magnetic clustering can be an issue that obfuscates the intrinsic properties of dilute magnetic systems \cite{Dietl14}.  Second, we find that Mn-Mn NNN dimers have relatively strong AF coupling.   To first order, such dimers form singlets that cannot participate in FM fluctuations.  However, the RKKY interaction that can drive FM order may couple dimers to isolated Mn ions, thereby inducing a magnetic moment on the dimer.  It will be interesting to examine dilute magnetic systems possessing long-range FM order to study the evolution of both the dimer excitations and FM fluctuations.  For example, FM-ordered (Bi$_{0.95}$Mn$_{0.05}$)$_{2}$Te$_{3}$ displays both dispersive collective magnetic modes and sharp excitations \cite{Vaknin19} that may reveal a unique coupling between magnetic moments and (topological) conduction electrons via the RKKY interaction.  

\section{Acknowledgments}
Work at the Ames Laboratory was supported by the U.S. Department of Energy (DOE), Basic Energy Sciences, Division of Materials Sciences and Engineering. Ames Laboratory is operated for the USDOE by Iowa State University under Contract No.~DE-AC02-07CH11358. A portion of this research used resources at the Spallation Neutron Source, a USDOE Office of Science User Facility operated by Oak Ridge National Laboratory.

\end{document}